# Repairing atomic vacancies in single-layer MoSe$_2$ field-effect transistor and its defect dynamics

Yuze Meng[1], Chongyi Ling[2], Run Xin[3], Peng Wang[4], You Song[5], Haijun Bu[1], Si Gao[4], Xuefeng Wang[3], Fengqi Song[1,*], Jinlan Wang[2,*], Xinran Wang[3,*], Baigeng Wang[1], Guanghou Wang[1]

[1] National Laboratory of Solid State Microstructures, Collaborative Innovation Center of Advanced Microstructures, and College of Physics, Nanjing University, Nanjing, 210093, P. R. China

[2] Department of Physics, Southeast University, Nanjing, P. R. China

[3] National Laboratory of Solid State Microstructures, Collaborative Innovation Center of Advanced Microstructures, and School of Electronic Science and Engineering, Nanjing University, Nanjing, 210093, P. R. China

[4] National Laboratory of Solid State Microstructures, Collaborative Innovation Center of Advanced Microstructures, and Department of Material Science and Engineering, Nanjing University, Nanjing, 210093, P. R. China

[5] State Key Laboratory of Coordination Chemistry, School of Chemistry and Chemical Engineering, Collaborative Innovation Center of Advanced Microstructures, Nanjing University, Nanjing, 210093, P. R. China

---

* Corresponding authors. songfengqi@nju.edu.cn; jlwang@seu.edu.cn; xrwang@nju.edu.cn. Meng Y., Ling C., Xin R. and Wang P. contribute equally. Fax:+86-25-83595535

**Abstract**

Here we repair the single-layer MoSe$_2$ field-effect transitors by the EDTA processing, after which the devices' room-temperature carrier mobility increases from 0.1 to over 70cm$^2$/Vs. The atomic dynamics is constructed by the combined study of the first-principle calculation, aberration-corrected transmission electron microscopy and Raman spectroscopy. Single/double Se vacancies are revealed originally, which cause some mid-gap impurity states and localize the device carriers. They are found repaired with the result of improved electronic transport. Such a picture is confirmed by a 1.5cm$^{-1}$ red shift in the Raman spectra.

**Text**

The new family of two-dimensional (2D) transition metal dichalcogenides materials, such as $MoS_2$, $WS_2$ and $WTe_2$, has attracted a lot of interest nowadays due to their excellent optical response[1-4], electronic control[4-6], layer-number controlled electronic structure [6-8] and extra valley degree of freedom[2, 9-12].etc[13]. Among them, the research attention on $MoSe_2$ is rising because of its high photoresponsivity and tunable exciton charge.[14, 15] It best matches the solar spectrum with its bandgap of 1.5eV. This leads to the potential applications in the optoelectronic devices, such as single-junction solar cells[16] and photoelechemical devices, based on single-layer $MoSe_2$ (SLM) with a direct bandgap and a even higher optical coupling strength.[14] Large SLM sheets with the dimension of over 100μm have been grown by chemical vapor deposition (CVD) method[17], based on which the vertical van der waals heterostructure devices are fabricated with the demonstration of rapid charge transfer in 50 $fs$[18]. The field-effect transistors based on both exfoliated and CVD grown SLM sheets have been fabricated, which on/off ratio reaches $10^{5\sim6}$ [17]. Lateral heterostructures $MoSe_2$-$WSe_2$, $MoSe_2$-$MoS_2$ have been grown, based on which the lateral p-n junctions were fabricated with a very long interlayer exciton lifetime of 1.8 $ns$[19]. Fast phototransistors have been demonstrated with the CVD-grown SLM sheets, whose response time is less than 25 $ms$.[20]

The defect engineering and its electronic control play a central role in the optimization of the 2D device performance[21-26]. The CVD-grown SLM sheets, which are expected to be the main source of future industrial-scale applications, shows various mobilities [17, 22, 27-30], which are far lower than the predicted extreme value of 240cm$^2$/Vs[31]. The defect dynamics of SLM sheets is still absent[32]. Here we report the improvement of the carrier mobility of the CVD-grown MoSe$_2$ from 0.1 cm$^2$/Vs to 20-70cm$^2$/Vs by using a simple chemical processing. The aberration-corrected high resolution transmission electron microscopy (HRTEM), Raman scattering and density functional theory (DFT) calculations form a complete defect dynamics of the Se vacancy repair, which is suggested the localization source of the transport carriers.

**Improved carrier mobility in SLM field-effect transistors after the processing**

The SLM sheets are grown by the CVD approach,[17] whose optical micrographs are shown in **Figure 1(a)**. We can see the length scale of the sheets are around 100μm. Our SLM sheets may reach 300-400μm. They are often of triangular shapes and very pale color contrast on the 500nm oxide Silicon wafers. Fig. 1(b) shows the atomic force microscopic image of our samples, where the thickness of 0.8nm is read according to the line profile as shown in the inset. Standard 2-probe electrodes are made to fabricate a field-effect transistor on selected monolayer sheets by the electron beam lithography technique as shown in the inset of Fig. 1(c). In Fig. 1(c), the measured current ($I_{sd}$) is plotted against the applied voltage ($V_{sd}$) for a typical sample measured at the room temperature (300K), where $I_{sd}$ is the source-drain current and $V_{sd}$ is the source-drain voltage. $I_{sd}$ increases from $10^{-13}$A at the gate voltage of -40V to the order of $10^{-9}$A at 40 V. It forms typical FET output curves, and also depicts SLM's n-type transport behavior. A Schottky contact is clear since the output line is not straight, based on which the contact barrier of 700 meV is extracted.[33] Such situation has been seen in CVD-grown MoS$_2$ devices.[21] The high resistance of up to $10^9 \Omega$ even at gate voltage of 40V and a low mobility of 0.1 cm$^2$/Vs reveal the localization of the transport carriers. Such strong localization has been seen in previous single layer MoS$_2$ devices since the defect interaction becomes stronger in reduced dimensions.[21, 22]

Processing the device by a drop of Ethylenediaminetetra acetic acid disodium salt solution (EDTA) improves the electronic transport as displayed in **Figure 2**. The processed sheet is kept in fuming cupboard for hours to dry off. We measure the transfer curve of the devices before and after the processing, which shows the data of log ($I_{sd}$) as a function of $V_g$ at room temperature (300K), where $V_g$ is the gate voltage. As shown in Fig.2(a), both the curves present the ON/OFF states at negative gate voltage and positive gate voltage respectively. At the given $V_{sd}$ of 0.5V, for the unprocessed sheet, $I_{sd}$ is in the order of $10^{-10}$ at gate voltage of -5V, and $I_{sd}$ increases by 1 orders to $10^{-9}$A at 20V. But after the EDTA processing, though the off current is still $10^{-10}$A, the on current comes to above $10^{-7}$A at the same gate voltage, which is 2-3 orders larger than the unprocessed ones. The ON/OFF ratio of the device may reach $10^4$ at 20V and we can expect a even higher ON/OFF ratio with a larger gate voltage. A linear trend governs in the positive part of the curve, by which we can calculate the field effect mobility by $\mu = L/W*(dG/dV_g)/C_i$,[22] where L is the distance between two probes, W is the sample width and $C_i$=6.96nFcm$^{-2}$ is the gate capacitance for 500nm SiO$_2$ dielectrics. $G=I_{sd}/V_{sd}$ where $V_{sd}$ is the source-drain voltage. We find that the unprocessed sample presents the mobility of 0.1 cm$^2$/Vs while the processed sample holds the mobility of 30cm$^2$/Vs. It reaches 70cm$^2$/Vs in some optimized device. It shows great mobility improvement by 2-3 order of magnitudes after the EDTA processing. Please note that the turn-on point of back gate voltage is above 0, which is the signature of the electronic dominance. We also estimate the carrier density of the device, which reveals the several time increase of the Fermi level after the processing.

Unlike MoS$_2$, MoSe$_2$ can be tuned to p-type transport when applying a negative back voltage[14]. We can't see the p-transport transition in the unprocessed SLM sheet even at the gate voltage of up to -150V. This might be due to the defect states. However, it can be achieved easily after the EDTA processing even at a very low back voltage. The EDTA processing may improve the hole mobility by 3 orders from below 0.01to 10 cm$^2$/Vs. Fig.2(b) shows $I_{sd}$ as a function of $V_g$ for the processed samples at room temperature. We can see from the $I_{sd}$-$V_g$ characteristics that $I_{sd}$ improves by 2 orders when the gate voltage changes from 0to -5V,which indicates the p-type transport. The hole mobility is calculated to be up to 10cm$^2$/Vs, similarly seen in some recent work[20]. The observation of p-type transport property in our SLM indicates the suppression of the localization.

We find the EDTA molecules improve the intrinsic properties of the MoSe$_2$ sheets rather than the contact between the metal electrodes and the sheets. In Fig. 2(c),we shows the data of Schottky barriers of a series of different samples at various gate voltages calculated by $U_s=k_BT/(-e)*ln(I_0/SA^*T^2)$[33], where $k_B$ is Boltzmann constant, $T$ is temperature, $e$ means electron charge, $S$ is area of probes, $A^*$is Richard constant $A^*=55A \cdot cm^2 \cdot K^2$ and $I_0$ is point of intersection of *log I* and y-axis in output characteristics. We can see the data of $U_s$ scatters around 0.6-0.8*V*, where no significant difference is found between processed and unprocessed samples. Therefore, it is convinced that the EDTA molecules improves the electronic localization and introduce n-type carriers, which dynamics forms the question of this work.

**The Se vacancies revealed by aberration-corrected HRTEM and its repair**

The aberration-corrected HRTEM is employed to study the defect states of the device before/after the EDTA processing as shown in **Figure 3**. Before the processing(Fig. 3(a)), the HRTEM reveals beautiful lattices in the monolayer sheet, indicating its genuine crystalline condition. Distinguished from their neighbors, some very dark features appear as marked by red arrows and magnified in Fig.3(a).After the EDTA processing, the HRTEM reveals only some scattered features with white spots among the triangular region. as marked by the green arrows shown in Fig. 3(b). Such HRTEM features have long been considered as the image of some atomic defects[22, 25, 34].We propose several atomic defect models and simulated their HRTEM images by the software Web-EMAPS[35]. We find that the SLM sheet with a double Se vacancy and the SLM sheet with an EDTA-filled Se vacancy best interpret the two defect images respectively. As seen in the simulated image in Fig. 3(c), the enhanced dark triangle corresponds to the vacant point in the build double vacancy model, which is similar to the observed features in Fig. 3(a). This means there are so much Se vacancies in the CVD-grown SLM sheet that some of them form double Se vacancies. The very low mobility is thus reasonable. Fig. 3(d) shows a model of EDTA-filled Se vacancy and its simulated HRTEM image. One may see a brighter triangular spot corresponding to the EDTA-filled vacancy. Similar features can be seen in Fig. 3(b) as stated above. This means the EDTA molecules repair the Se vacancies significantly during the processing.

This picture is confirmed by the Raman measurement as shown in **Figure 4(a)**, which finds the Raman feature appear at 240.2cm$^{-1}$. It is believed to be the characteristic peak of MoSe$_2$[36] and shifted to 238.7 cm$^{-1}$ after the processing. This is consistent with the repair model. We calculate the phonon dispersion curves of unprocessed and processed MoSe$_2$ are shown in fig. 4(b), in which the Raman active modes ($A_1\check{}$) at Γ point were presented. It can be vividly observed that the $A_1\check{}$ modes shift to lower frequency at Γ point after the processing, which is consistent with our experimental results. A simple picture is the Raman shift characterizes the phonon scattering, which frequency is related to the mass of the atomic chains. The repairing of the Se vacancy tunes the resonant frequency and leads to the observed Raman shift.

**The repair dynamics from the first-principle calculations**

The DFT calculations complete the story by a Se vacancy repair scenario. The model is a supercell consist of 4*4 SLM unit cell with a single Se vacancy as shown in **Figure 5(c)**. The attachment of the EDTA molecule is found to connect by the

COO- bond and the form tow Mo-O bonds with the Mo atoms around the Se vacancy. In fact, various models of EDAT adsorption have been tried, including a pristine EDTA molecule adsorbed on the surface of pure $MoSe_2$ or $MoSe_2$ with a Se vacancy, where most of them present a very weak binding strength. The calculated electronic structure was shown in Fig. 5(a), in which some flat bands arise in the gap (denoted by red curves). These localized bands are contributed by the three Mo atoms around the Se vacancy according to our calculations about partial charge density distribution, as shown in the inset of Fig. 5(a). This means because the Se vacancies exist in the SLM sheet, the electrons are trapped around the central atoms as shown in the yellow circles with the result of a strong localization. After the processing of EDTA molecule, the flat impurity bands are broadened, which means these bands become much more extended. The Fermi level is found lifted a bit after the repair. We are therefore convinced that the repair of the Se vacancy leads to the improvement of the carrier mobility and the increased Fermi level.

**Conclusion**

In conclusion, we have improved the mobility of SLM devices by 3 orders by a simple EDTA processing. Both the electron and hole mobility are improved. HRTEM and Raman studies reveal the repair of the single/double Se vacancies, which localize the electronic transport in the single layer devices. The scenario is completed by a density functional theory calculation. This work will further pave the advance of the SLM applications in practical optoelectronic uses.

**Methods**

**The sample growth and processing.** We used CVD to synthesize the SLM sheets, where Selenium pellets (Se) (Alfa Aesar 99.99%) and molybdenum oxide ($MoO_3$) (Alfa Aesar, 99.5%) power were used as Se and Mo precursors. They were placed in two different quartz boats and Se power was on the windward side. A clean Si wafer with a 500nm (or 300nm) $SiO_2$ layer was placed face down on the Mo boat. The Mo boat was then put at the center of the heating furnace. During the whole reaction, Ar/$H_2$ gas flux was kept at 50sccm as the carrier gas. The furnace temperature was held at 30°C for 1 hour in order to remove $O_2$. Then we raised the furnace to 720°C with a heating ramp of 24°C/min. After half an hour, the furnace was kept at 720°C for 12 minutes and cooled down to room temperature. The whole reaction was carried on ambient pressure. The EDTA solution is dropped on the device.

**The Characterization and HRTEM simulation.** The HRTEM was carried out in a FEI TITAN CUBED 60-300 machine. The acceleration voltage of 60V is used at the magnification of 10000000X. The Raman spectroscopy was carried out in an NT-MDT NANOFI NEDR-300. All the transport measurement was carried out in a home-made system with a Keithley 4200 and a Keithley 6430. The HRTEM simulation was carried out in a standard electron microscopy application software developed by Zuo et al, where the supercell is the input structure as shown in Fig. 4(c,d). The beam energy is 60kV. A series of defocus parameters are tried to obtain the correct images.

**The density functional theory calculation.** The DFT calculations are performed using the pseudopotential plane-wave method with projected augmented wave[37] potentials and Perdew-Burke-Ernzerhof-type generalized gradient approximation (GGA) [38] for exchange-correlation functional, as implemented in the Vienna *ab initio* simulation package (VASP) [39]. The plane-wave energy cutoff is set to be 400 eV.

**Acknowledgement**

We gratefully acknowledge the financial support of the National Key Projects for Basic Research of China (Grant Nos: 2013CB922100,2011CB922103), the National Natural Science Foundation of China (Grant Nos: 91421109, 11134005, 11522432，11474147, 21571097 and 11274003), the PAPD project, the Natural Science Foundation of Jiangsu Province (Grant BK20130054), and the Fundamental Research Funds for the Central Universities. We would also like to acknowledge the helpful assistance of the Nanofabrication and Characterization Center at the Physics College of Nanjing University.

# Figure Caption

**Figure 1. The original performance of the SLM device.** (a) Optical photograph of monolayer $MoSe_2$; (b). AFM picture of monolayer $MoSe_2$; (c). $I_{sd}$-$V_{sd}$ characteristics of CVD-synthesized monolayer $MoSe_2$ at gate voltages of -20,0,20,40V.

**Figure 2. Repairded SLM device with a 3-order higher mobility.** (a). $I_{sd}$-$V_g$ curves of CVD-synthesized monolayer $MoSe_2$ before (black) and after (red) the EDTA processing; (b). $I_{sd}$-$V_g$ curves of another sample processed by EDTA, shows the improved hole mobility of SLM sheet; (c). The Schottky barrier values of five different samples, black means unprocessed and red means processed respectively.

**Figure 3. The atomic vacancy dynamics revealed by HRTEM.** (a). the HRTEM image of unprocessed SLM sheet, red arrows show the double Se vacancy; (b).HRTEM photograph of processed monolayer $MoSe_2$,rad arrows show the double Se vacancy, green arrows show the double Se vacancy processed by one side with EDTA; (c).double Se vacancy in monolayer $MoSe_2$ and its simulated TEM photograph;(d)double Se vacancy processed by one side with EDTA-2Na in monolayer $MoSe_2$ and its simulated TEM photograph;

**Figure 4. The vacancy repair confirmed by Raman spectroscopy.** (a)Raman spectra of unprocessed (red) and processed(black) SLM sheet; (b) The Calculated phonon dispersion curves of unprocessed(red) and processed (black) SLM sheet respectively.

**Figure 5. The first-principle calculation**. The band structures and electron density of single-Se-vacancy SLM without (a) and with (b) the adsorption of EDTA. yellow: electron density; (c)EDTA absorbed on the SLM. red: O ; white: H ; grey: C ; blue: N.

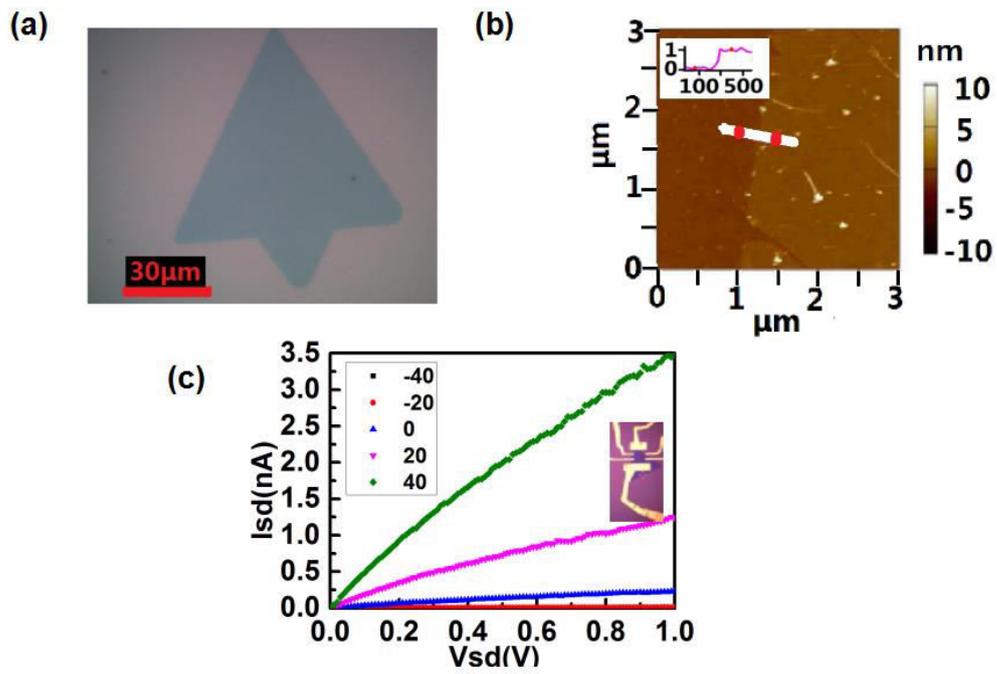

**Figure 1 Meng et al**

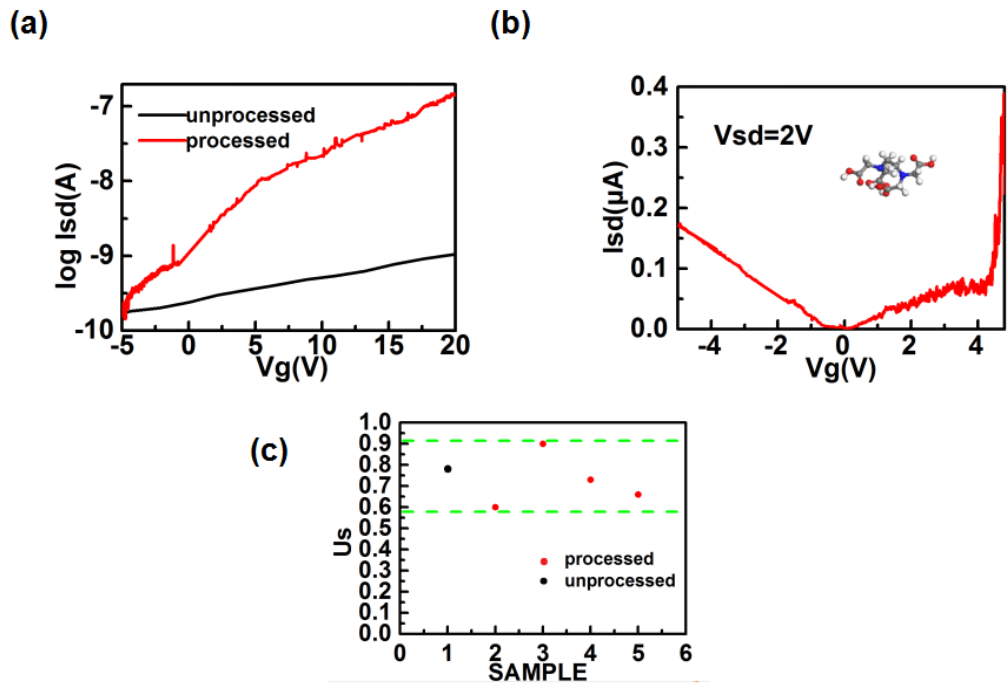

Figure 2 Meng et al

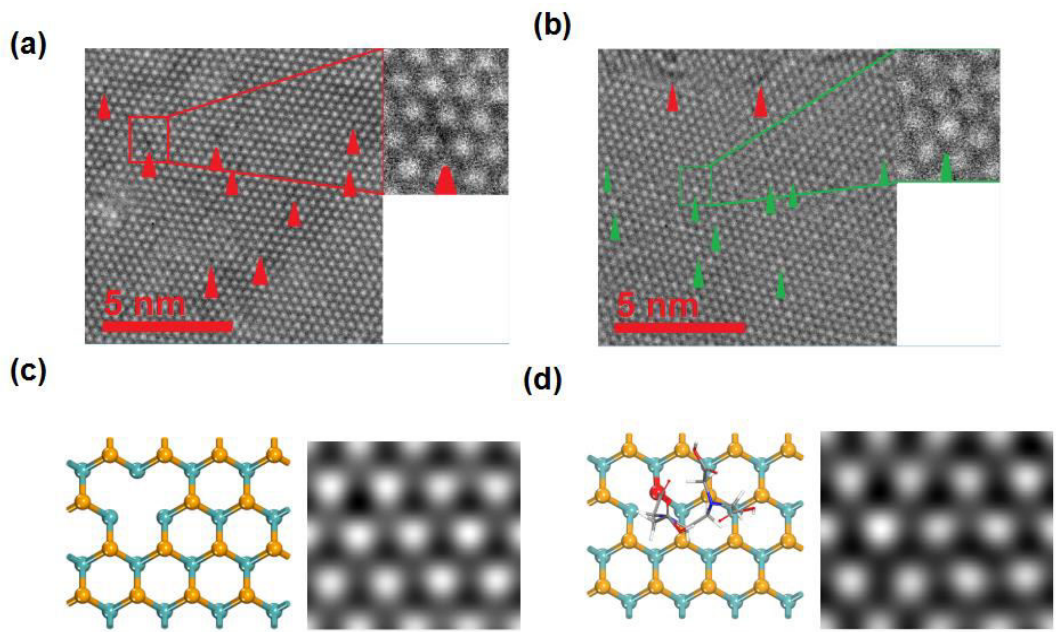

**Figure 3 Meng et al**

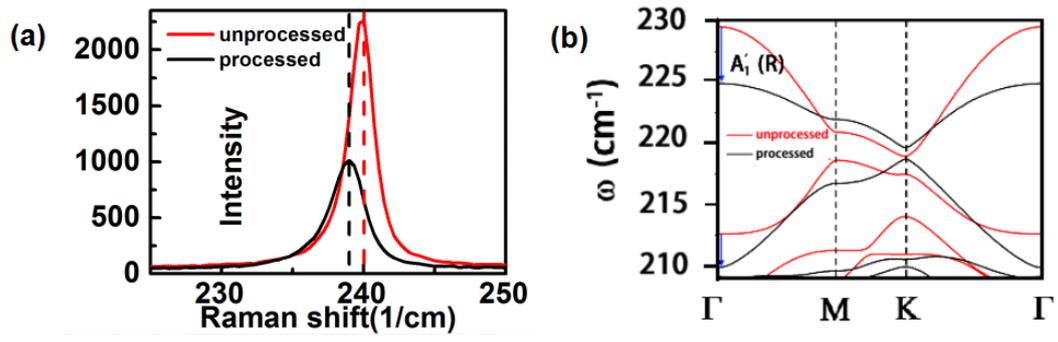

**Figure 4 Meng et al**

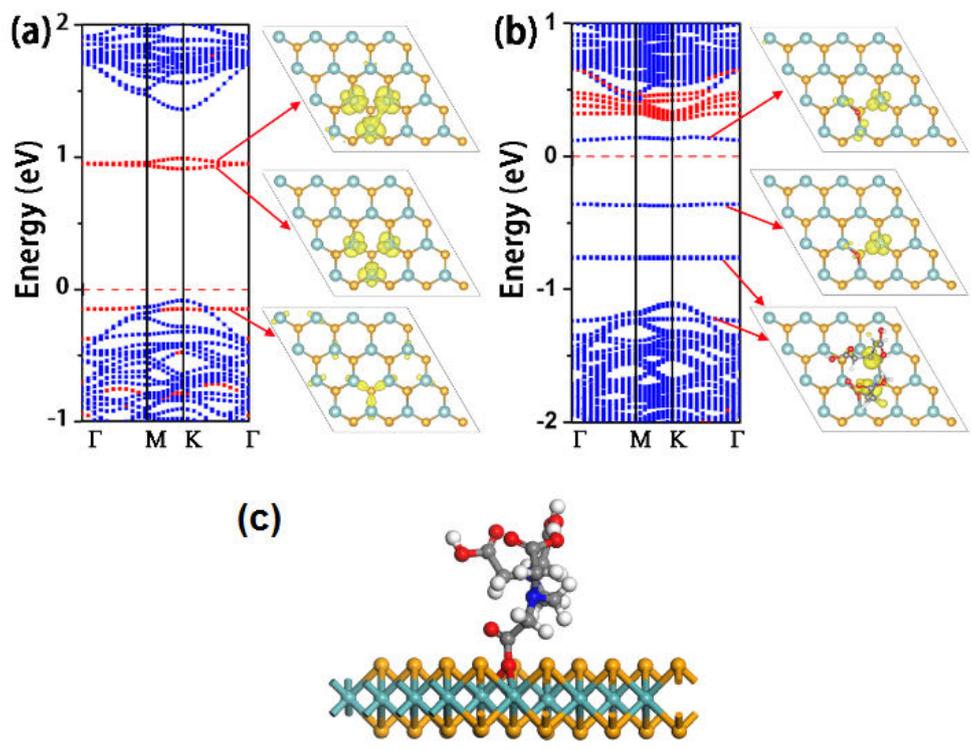

**Figure 5 Meng et al**

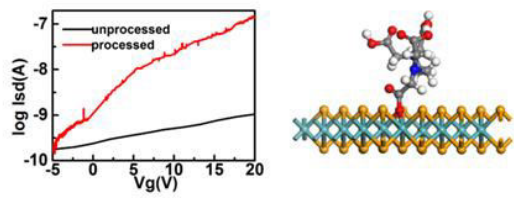

**The TOC figure**

The EDTA molecule repairs the localization source, Se vacancy, in the MoSe$_2$ single-layer field-effect transitors and improves its mobility by 3 orders.